\documentclass[prb,fleqn,12pt,showpacs,superscriptaddress]{revtex4}
\usepackage{epsfig}
\usepackage{graphicx}
\begin{document}
\title{Spin waves in the AF state of the $t$-$t'$ Hubbard model on the fcc lattice: 
competing interactions, frustration, and instabilities}
\author{Avinash Singh}
\email{avinas@iitk.ac.in}
\affiliation{Institute Laue Langevin (ILL), 71 avenue des Martyrs - CS 20156 - 38042 GRENOBLE CEDEX 9, France} 
\affiliation{Department of Physics, Indian Institute of Technology Kanpur - 208016, India}
\author{Shubhajyoti Mohapatra}
\affiliation{Department of Physics, Indian Institute of Technology Kanpur - 208016, India}
\author{Timothy Ziman}
\affiliation{Institute Laue Langevin (ILL), 71 avenue des Martyrs - CS 20156 - 38042 GRENOBLE CEDEX 9, France} 
\author{Tapan Chatterji} 
\affiliation{Institute Laue Langevin (ILL), 71 avenue des Martyrs - CS 20156 - 38042 GRENOBLE CEDEX 9, France} 
\date{\today} 
\begin{abstract}
Spin waves in the type-III ordered antiferromagnetic state of the frustrated $t$-$t'$ Hubbard model on the fcc lattice are calculated to investigate finite-$U$-induced competing interaction and frustration effects on magnetic excitations and instabilities. Particularly strong competing interactions generated due to interplay of fcc lattice geometry and magnetic  order result in significant spin wave softening. The calculated spin wave dispersion is found to be in qualitative agreement with the measured spin wave dispersion in the pyrite mineral $\rm Mn S_2$ obtained from inelastic neutron scattering experiments. Instabilities to other magnetic orders (type I, type II, spiral, non-collinear), as signalled by spin wave energies turning negative, are also discussed. 


\end{abstract}


\maketitle

\newpage

\section{Introduction}

Frustrated magnetism continues to be of considerable current interest due to the rich possibility of new states and properties of matter.\cite{lacroix_2011} While Kagome and triangular lattices have been widely studied, frustration in the face-centred-cubic (fcc) lattice has received much less attention, particularly within itinerant electron models. Antiferromagnetic (AF) orders in fcc materials range from type-III and type-I in the 1:2 compounds ${\rm Mn S_2}$ and ${\rm Mn Te_2}$,\cite{hastings_1959} to type-II in the 1:1 compound ${\rm MnO}$.\cite{bloch_1974} Magnetic frustration and pressure-induced metal-insulator transition are exhibited by a variety of complex compounds having effective fcc magnetic lattice such as alkali fullerides $\rm A_3 C_{60}$ (A = K, Rb, Cs),\cite{capone_2009,ganin_2010} cluster compounds $\rm Ga Ta_4 Se_8 $, $\rm Ga Nb_4 Se_8 $,\cite{guiot_2013} and `B site ordered' double perovskites.\cite{aczel_2013} 

Magnetic frustration in an itinerant electron model has additional features besides the usual geometric frustration effect in the ideal Heisenberg model with nearest-neighbor (NN) two-spin interaction where the interplay of magnetic interactions and lattice geometry results in frustrated spins. This is effectively illustrated by the case of the $120^{\circ}$ ordered AF state of the Hubbard model on a triangular lattice. While at large $U$, spin wave dispersion in the random phase approximation (RPA) exactly matches with the corresponding result for the Heisenberg model with $J=4t^2/U$, extended-range effective spin couplings generated at finite $U$ result in strong zone-boundary spin wave  softening and even magnetic instability with decreasing $U$, highlighting the finite-$U$-induced competing interaction and frustration effect in an itinerant electron system.\cite{as_2005} The cyclic ring-exchange four-spin term $({\bf S}_i \times {\bf S}_j).({\bf S}_k \times {\bf S}_l)$, generated in the square-lattice Hubbard model at next-to-leading order in the $t/U$ expansion arising from coherent motion of electrons beyond NN sites, illustrates that higher-spin couplings are also generated besides extended-range two-spin interactions.\cite{coldea_2001} 

The itinerant electron approach also directly connects magnetic frustration and spin density wave (SDW) band gap. The same hopping terms between parallel spins which are responsible for competing interactions, also result in band broadening which strongly reduces the SDW gap and renders the system more susceptible to metal-insulator transition with decreasing $U/t$. SDW band gap reduction, band overlap, and first-order metal-insulator transition with decreasing $U/t$ have been studied in the frustrated square- and triangular-lattice antiferromagnets due to electron self-energy correction calculated in the self consistent Born approximation (SCBA).\cite{asingh_2004+5}

Within the $t$-$t'$ Hubbard model on the fcc lattice, ground-state magnetic phase diagram and related metal-insulator transition have been investigated very recently using the slave-boson approach by minimizing the ground state thermodynamical potential with respect to the spiral state wave vector.\cite{timirgazin_2016} The magnetic phase diagrams for different band fillings (fixed $t'$) as well as for different $t'$ (half filling) were obtained, showing a variety of magnetic phases and transitions. However, finite-$U$-induced competing interaction and frustration effects on spin waves have not been investigated for the fcc-lattice AF state of the Hubbard model. Study of magnetic excitations should be of particular interest for type-III order in view of the measured spin wave dispersion in $\rm MnS_2$, obtained from inelastic neutron scattering studies of the naturally occuring pyrite-structured mineral hauerite.\cite{tapan_mns2}

\begin{figure}
\vspace*{0mm}
\hspace*{0mm}
\psfig{figure=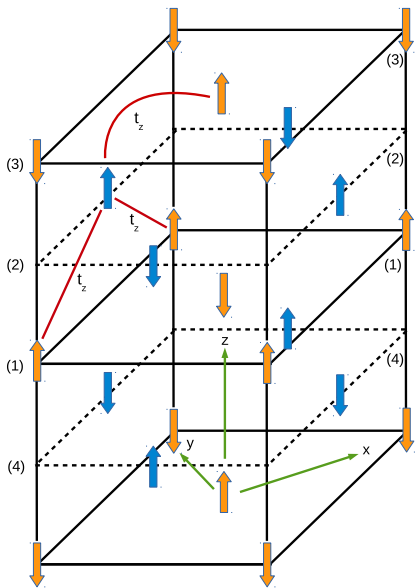,angle=0,width=60mm}
\caption{Type-III AF order on the fcc lattice. Planes shown in solid and dashed lines with spins in red and blue indicate the two identical fcc sublattices. The layers along the $z$ direction in the sequence $\alpha \alpha' \beta \beta' \alpha ...$ (labeled as $12341...$) have planar $(\pi,\pi)$ magnetic order along the $x$ and $y$ directions shown. Parallel spins connected by NN hopping in different fcc sublattices ($t_z$) reflect the strong inherent magnetic frustration.}
\end{figure}
 
\section{AF orders on the fcc lattice}

Neutron scattering studies of the AF structures of ${\rm MnS_2}$, ${\rm MnSe_2}$, and ${\rm MnTe_2}$ have shown orderings of the ``third" kind for the disulphide, of the ``first" kind for the ditelluride, and an intermediate arrangement for the diselenide.\cite{hastings_1959} The magnetic structure has been established as collinear for ${\rm MnS_2}$.\cite{tapan_1991} While a planar $(\pi,\pi)$ order ($xy$ plane in Fig. 1) of nearest-neighbour (NN) spins is common to all three, it is the order in the perpendicular ($z$) direction involving next-nearest-neighbour (NNN) spins which distinguishes the three cases. The order is AF for the disulphide (smallest lattice parameter 6.097 \AA) and F for the ditelluride (largest lattice parameter 6.943 \AA), whereas with intermediate lattice parameter 6.417 \AA, the diselenide exhibits the intermediate arrangement. These structures may therefore be regarded as different interlayer stackings of the AF ordered layers. As inferred from the relatively low $T_{\rm N}$ values, the weak magnetic couplings between the Mn spins are due to the relatively large lattice parameters in these high-spin ($S$=5/2) systems. AF order of the ``second" kind is realized in the 1:1 compound $\rm Mn O$ (much smaller lattice parameter 4.447 \AA), which has planar $(\pi,0)$ order instead.  


These fcc lattice antiferromagnets exhibit several unusual magnetic properties. For $\rm MnS_2$, the magnetic phase transition at $T_{\rm N}$=48 K is of first order.\cite{hastings_1976,tapan_1984,westrum_1970} By using very high resolution synchrotron x-ray diffraction techniques, a pseudo-tetragonal distortion was detected below the magnetic ordering temperature ($c/a$ ratio 1.0006), indicating coupling between magnetic and lattice degrees of freedom.\cite{kimber_2015} Giant pressure-induced volume collapse accompanied by high spin ($S=5/2$) to low spin ($S=1/2$) transition involving interplay between crystal field splitting and Hund's rule coupling has been observed in this pyrite 
mineral.\cite{kimber_2014} The measured N\'{e}el temperature of $\rm MnTe_2$ has been found to show unusually large pressure dependence of 12K/GPa, giving rise to large violation of Bloch's rule.\cite{tapan_2015} Based on IR reflection measurements at room temperature, $\rm MnTe_2$ appears to undergo pressure-induced semiconductor-metal transition in the pressure range of 8-25 GPa.\cite{mita_2008}

Quantum Monte Carlo simulations in a NN classical Heisenberg AF on the fcc lattice have confirmed a first order transition to a collinear type-I AF structure due to an ``order by disorder" effect.\cite{gvozdikova_2005} A first-order transition driven by thermal fluctuations has been suggested by the absence of stable fixed points within the renormalization group approach.\cite{qmc1,qmc2} As an illustration of low-temperature thermal fluctuations selecting collinear states through the ``order by disorder" effect,\cite{henley_1987} short wavelength thermal fluctuations lead to an effective biquadratic exchange $-({\bf S}_i . {\bf S}_j)^2$ between neighboring spins,\cite{canals_2004} which favours collinear spin arrangement. 

Strong geometric frustration is inherent in these fcc-lattice antiferromagnets. Unlike the weakly frustrated square-lattice AF, it is the strong NN AF bonds in neighboring layers which are frustrated in the fcc lattice. Also, within the localized-spin picture, type-III (type-I) order on the fcc lattice is stabilized for AF (F) sign of the second-neighbor interaction,\cite{henley_1987} as expected from Fig. 1. Competing interactions between neighboring layers of same and different fcc sublattices also allows for the spiral spin structure. In this paper, we will show through a spin wave stability analysis that strong finite-$U$-induced competing interaction and frustration effects in the fcc lattice result in significant additional spin wave softening (besides the usual geometric frustration effect), which considerably enriches the competition between different AF orders in the $t$-$t'$ Hubbard model. 


\section{$t$-$t'$ Hubbard model}

We consider the $t$-$t'$ Hubbard model on the fcc lattice:
\begin{equation}
H = -t \sum_{\langle i,j \rangle,\sigma} a_{i\sigma} ^\dagger a_{j\sigma}
- t' \sum_{\langle \langle i,j \rangle \rangle,\sigma} a_{i\sigma} ^\dagger a_{j\sigma}
+ U \sum_i n_{i\uparrow} n_{i\downarrow} 
\end{equation}
where $t$ and $t'$ are the nearest- and next-nearest-neighbour hopping terms, respectively, and $U$ is the on-site Coulomb interaction. In order to identify the role of fcc lattice in  magnetic frustration, we will employ the interlayer NN hopping terms (shown as $t_z$ in Fig. 1) as control. For $t_z$=0, the two fcc sublattices are completely decoupled, while they are coupled for $t_z$=$t$ (cubic case). In the following, we will set $t$=1 as the energy scale. 

Type-III order on the fcc lattice is shown in Fig. 1. Alternating layers along $z$ direction, shown as planes in solid and dashed lines with spins in red and blue, constitute two identical fcc sublattices. The type-III order is characterized by $(\pi,\pi)$ magnetic order in each layer, with layers within same fcc sublattice stacked antiferromagnetically in the $z$ direction. The NNN hopping term $t'$ provides the weak AF interlayer coupling required for stabilizing type-III order. Within the equivalent localized spin model (large $U$ limit), the NNN spin coupling $J'$ connects spins only within same fcc sublattice (the weak AF interlayer coupling), whereas the NN spin coupling $J$ connects spins in different fcc sublattices as well. These latter interactions are fully frustrated, and the relative magnetic orientation between the two fcc sublattices can therefore be arbitrary in the classical ground state. 

Corresponding to the type-III order, we consider the interaction term (Eq. 1) in the Hartree-Fock (HF) approximation with local magnetization taken along the $z$ direction and staggered field $\mp \sigma \Delta$ on the two magnetic sublattices A and B. In a composite four-layer $\otimes$ two-sublattice basis corresponding to the magnetic order, we obtain the $8\times 8$ Hamiltonian matrix:

\begin{equation}
H_{\rm HF} ^\sigma (\bf k) = \left [ 
\begin{array}{cccccccc}
\varepsilon_{kxy}' & \varepsilon_{kxy} & \varepsilon_{kzxy} & \varepsilon_{kz\overline{xy}} & 0 & \varepsilon_{kz}' & \varepsilon_{kz\overline{xy}}^* & \varepsilon_{kzxy}^*  \\
  & \varepsilon_{kxy}' & \varepsilon_{kz\overline{xy}} & \varepsilon_{kzxy} & \varepsilon_{kz}' & 0 & \varepsilon_{kzxy}^* & \varepsilon_{kz\overline{xy}}^*  \\
  &  &  \varepsilon_{kxy}' & \varepsilon_{kxy} & \varepsilon_{kz\overline{xy}} & \varepsilon_{kzxy} & 0 & \varepsilon_{kz}'  \\
  &  &  & \varepsilon_{kxy}' & \varepsilon_{kzxy} & \varepsilon_{kz\overline{xy}} & \varepsilon_{kz}' & 0  \\
  &  &  &  & \varepsilon_{kxy}' & \varepsilon_{kxy} & \varepsilon_{kzxy} & \varepsilon_{kz\overline{xy}}  \\ 
  &  &  &  &  & \varepsilon_{kxy}' & \varepsilon_{kz\overline{xy}} & \varepsilon_{kzxy}  \\ 
  &  &  &  &  &  & \varepsilon_{kxy}' & \varepsilon_{kxy}  \\ 
  &  &  &  &  &  &  & \varepsilon_{kxy}' \\ 
\end{array} \right ] \mbox{$\mp \sigma \Delta$}
\end{equation} 
where the band terms corresponding to NN and NNN hoppings in the planar ($xy)$ and perpendicular ($z$) directions are given by:
\begin{eqnarray}
\varepsilon_{kxy}' & = & -4t' \cos k_x \cos k_y \\ \nonumber
\varepsilon_{kxy} & = & -2t (\cos k_x + \cos k_y) \\ \nonumber
\varepsilon_{kz}' & = & -2t' \cos k_z  \\ \nonumber
\varepsilon_{kzxy} & = & -2t_z e^{i k_z /2} \cos \left (\frac{k_x + k_y}{2} \right )  \\ \nonumber
\varepsilon_{kz\overline{xy}} & = & -2t_z e^{i k_z /2} \cos \left (\frac{k_x - k_y}{2} \right ) 
\end{eqnarray}
Here $\Delta = mU/2$ is the staggered field in terms of the sublattice magnetization:
\begin{equation}
m (\Delta) 
= (n_\uparrow ^A - n_\downarrow ^A)(\Delta) 
= (n_\downarrow ^B - n_\uparrow ^B)(\Delta) 
= (n_\uparrow ^A - n_\uparrow ^B)(\Delta)
\end{equation}
which is determined self-consistently from the electronic densities calculated from $H_{\rm HF} ^\sigma (\bf k)$ for the two spins $\sigma$=$\uparrow,\downarrow$ on the two magnetic sublattices A and B. In practice, it is easier to choose $\Delta$ and determine $U$ from the calculated sublattice magnetization $m(\Delta)$. In the large $U$ limit, $2\Delta \approx U$ as $m \rightarrow 1$. We will consider only the half-filled case ($n=1$) with Fermi energy in the AF band gap. 
Note that our coordinate axes ($x-y$) are rotated by $\pi/4$ with respect to the cubic planar axes, with lattice parameter $a/\sqrt{2}$ for the corresponding square lattice. Therefore $k_x,k_y$ and $k_z$ are in units of $\sqrt{2}/a$ and $1/a$, respectively, in terms of the cubic lattice parameter $a$. 

\begin{figure}
\vspace*{0mm}
\hspace*{0mm}
\psfig{figure=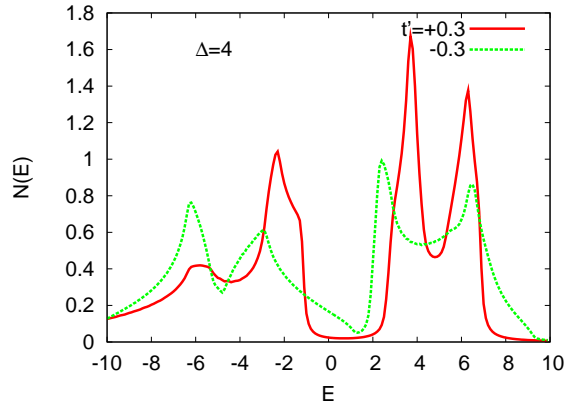,angle=-90,width=80mm}
\caption{The HF level electronic DOS in the AF state, showing strongly reduced SDW band gap due to frustration compared to the unfrustrated band gap ($2\Delta$), and strongly asymmetric behaviour with respect to sign of NNN hopping term $t'$. Here $U$=9.2 (9.7) for $t'=+(-)0.3$.}
\end{figure}

\section{AF state electronic density of states}

The AF state electronic density of states (DOS) shows strongly asymmetric behaviour with respect to sign of $t'$ (Fig. 2). For positive $t'$, the SDW band gap is more robust, and the AF insulator state survives even for relatively lower U values. DOS structure is similar to that for the planar $t$-$t'$ model, except that the fcc hopping term $t_z$ further splits the two SDW bands. The DOS drops off sharply at both band edges. Electronic self energy correction, as incorporated at the SCBA level, will further reduce the band gap, resulting in band overlap with decreasing $U/t$.

On the other hand, for negative $t'$, the SDW band gap is significantly reduced due to band broadening, and the AF insulator state requires higher $U$ values. DOS structure is different from the planar case, indicating more three dimensional band structure effect due to the $t_z$ hopping term. The DOS does not fall abruptly as in the previous case but has broad tail for the lower band, indicating possibility of metallic AF state surviving even after weak band overlap, and similarly for small hole doping. 

\section{Spin wave excitations} 

We consider the spin wave propagator: 
\begin{equation}
\chi^{-+}({\bf q},\omega) = \int dt \sum_{i} e^{i\omega(t-t')} e^{-{\bf q}.({\bf r}_i - {\bf r}_j)} \langle \Psi_0 | {\rm T} [ S_i ^{-}(t) S_j ^{+}(t') ] | \Psi_0 \rangle
\end{equation}
obtained from expectation value of the time-ordered product of  transverse spin operators $S_i ^-$ and $S_j ^+$ at lattice sites $i$ and $j$ in the AF ground state $|\Psi_0 \rangle$. In the random phase approximation (RPA), the spin wave propagator can be written in the composite basis as:
\begin{equation}
[\chi^{-+}({\bf q},\omega)] = \frac{[\chi^0({\bf q},\omega)]} {{\bf 1} - U [\chi^0({\bf q},\omega)]}
\end{equation}
where $[\chi^0]$ is the bare particle-hole propagator matrix obtained by integrating out fermions in the broken-symmetry state. In terms of the energy eigenfunctions $\phi_{\bf k}$ and eigenvalues $E_{\bf k}$ of the Hamiltonian matrix $H_{\rm HF}^{\sigma} ({\bf k})$,
\begin{eqnarray}
[\chi^{0}({\bf q},\omega)]_{ss'} &=& i \int \frac{d\omega^{\prime}}{2\pi} \sum_{\bf k'} \left[G_{0}^{\uparrow}({\bf k'},\omega^{\prime}) \right]_{ss'} \left[G_{0}^{\downarrow}({\bf k'-q},\omega^{\prime} - \omega) \right]_{s's} \nonumber \\ &=& \sum_{{\bf k'},m,n} \left[ \frac{\phi^{\uparrow s}_{{\bf k'}\,m} \phi^{\uparrow s' *}_{{\bf k'}\,m} \phi^{\downarrow s'}_{{\bf k'-q}\,n} \phi^{\downarrow s *}_{{\bf k'-q}\,n}} {E^{+}_{{\bf k'-q}\downarrow n} - E^{-}_{{\bf k'}\uparrow m} + \omega -i\eta} 
+ \frac{\phi^{\uparrow s}_{{\bf k'}\,m} \phi^{\uparrow s' *}_{{\bf k'}\,m} \phi^{\downarrow s'}_{{\bf k'-q}\,n} \phi^{\downarrow s *}_{{\bf k'-q}\,n}}{E^{+}_{{\bf k'}\uparrow m} - E^{-}_{{\bf k'-q}\downarrow n} - \omega - i\eta} \right] 
\end{eqnarray}
Here $s,s'$ refer to indices in the composite four-layer, two-sublattice basis, $m$,$n$ indicate the eigenvalue branches, and + (-) refer to particle (hole) energies above (below) the Fermi energy. By diagonalizing the $[\chi^0({\bf q},\omega)]$ matrix, spin wave energies are obtained from solutions of 1-$U\lambda_{\bf q} ^l (\omega)$=0 representing poles of Eq. (6). Corresponding to the four-layer basis, there are four spin-wave branches. 

\begin{figure}
\vspace*{0mm}
\hspace*{0mm}
\psfig{figure=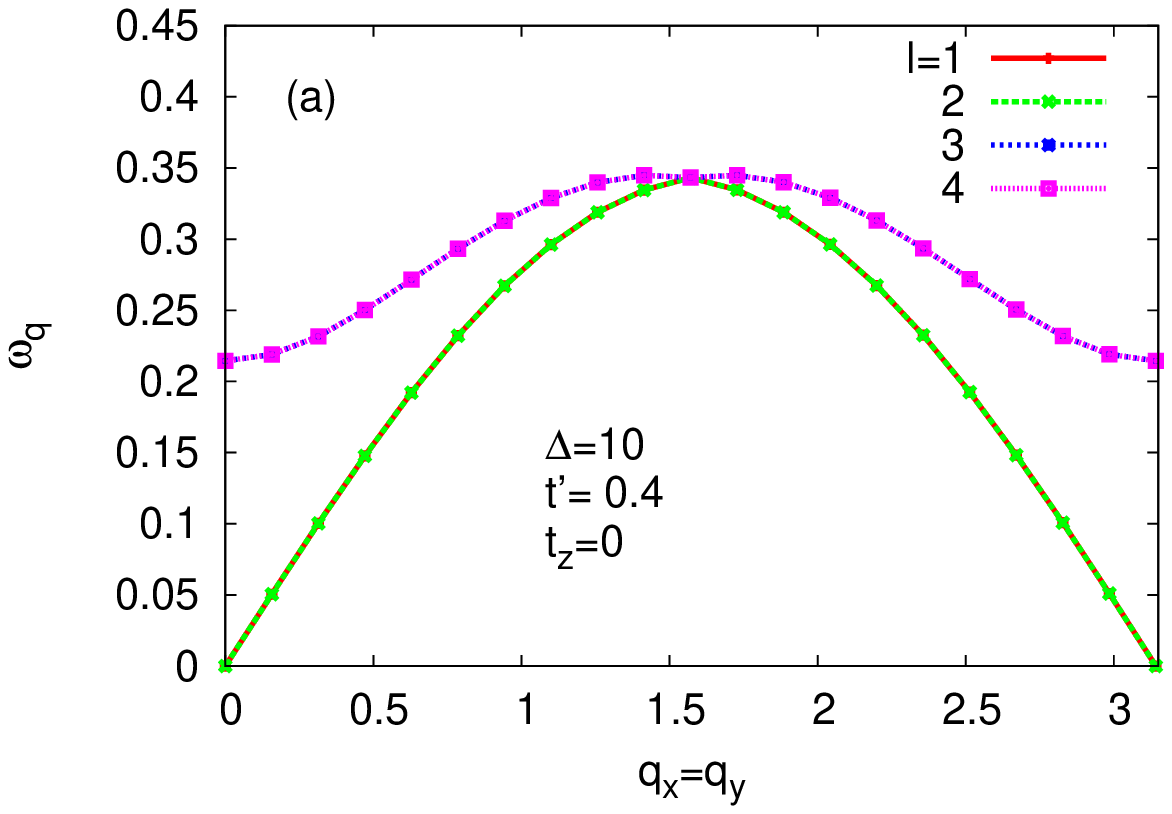,angle=0,width=70mm}
\psfig{figure=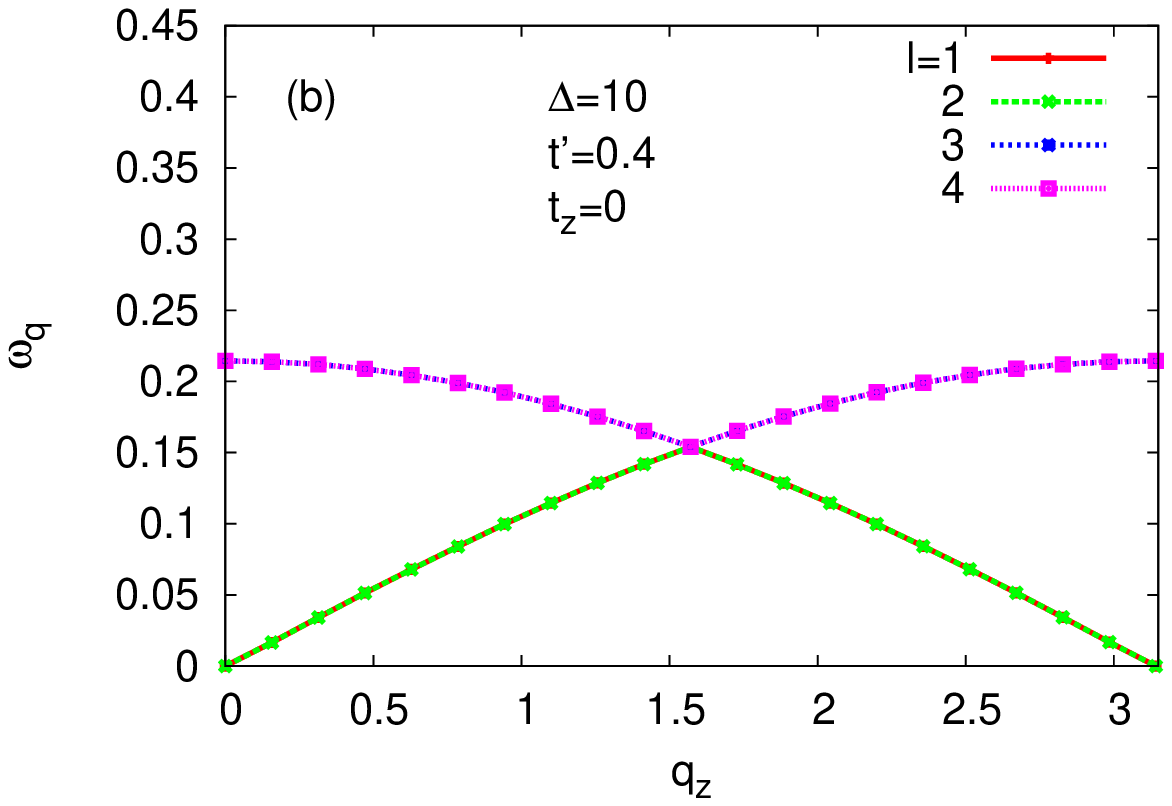,angle=0,width=70mm}
\ \\
\psfig{figure=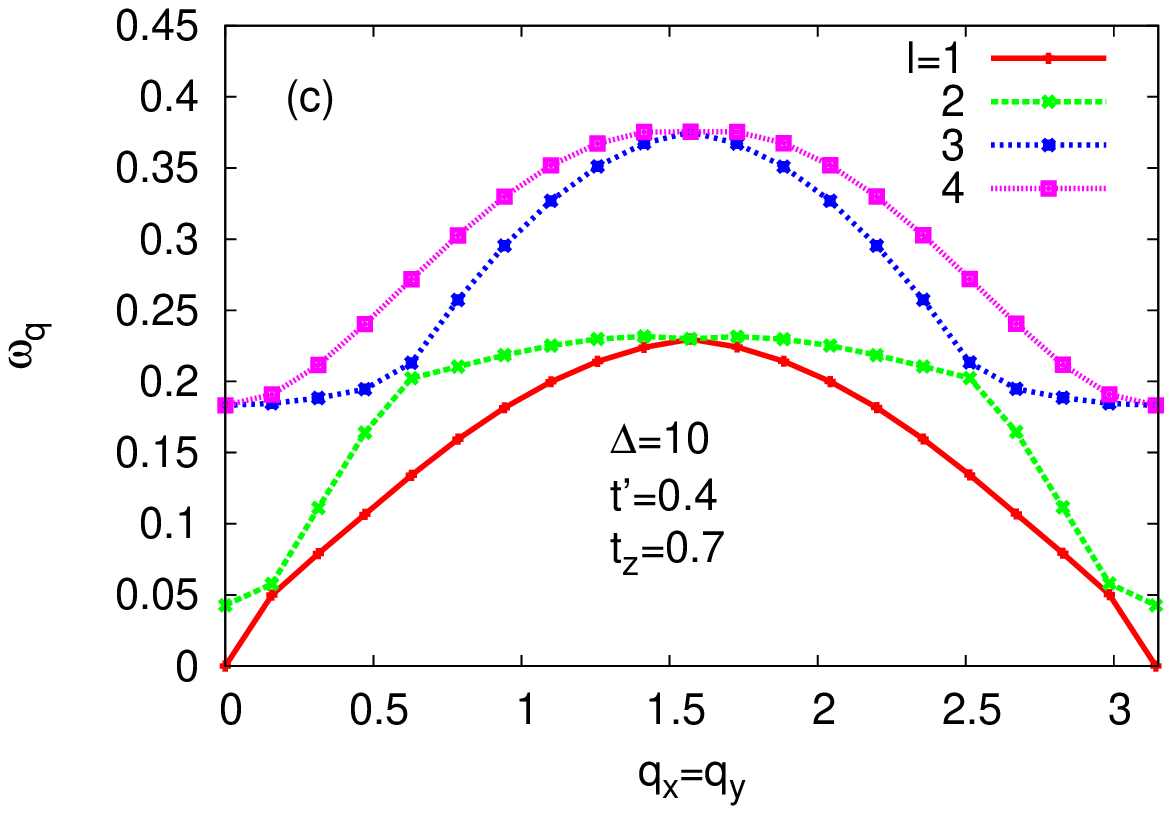,angle=0,width=70mm}
\psfig{figure=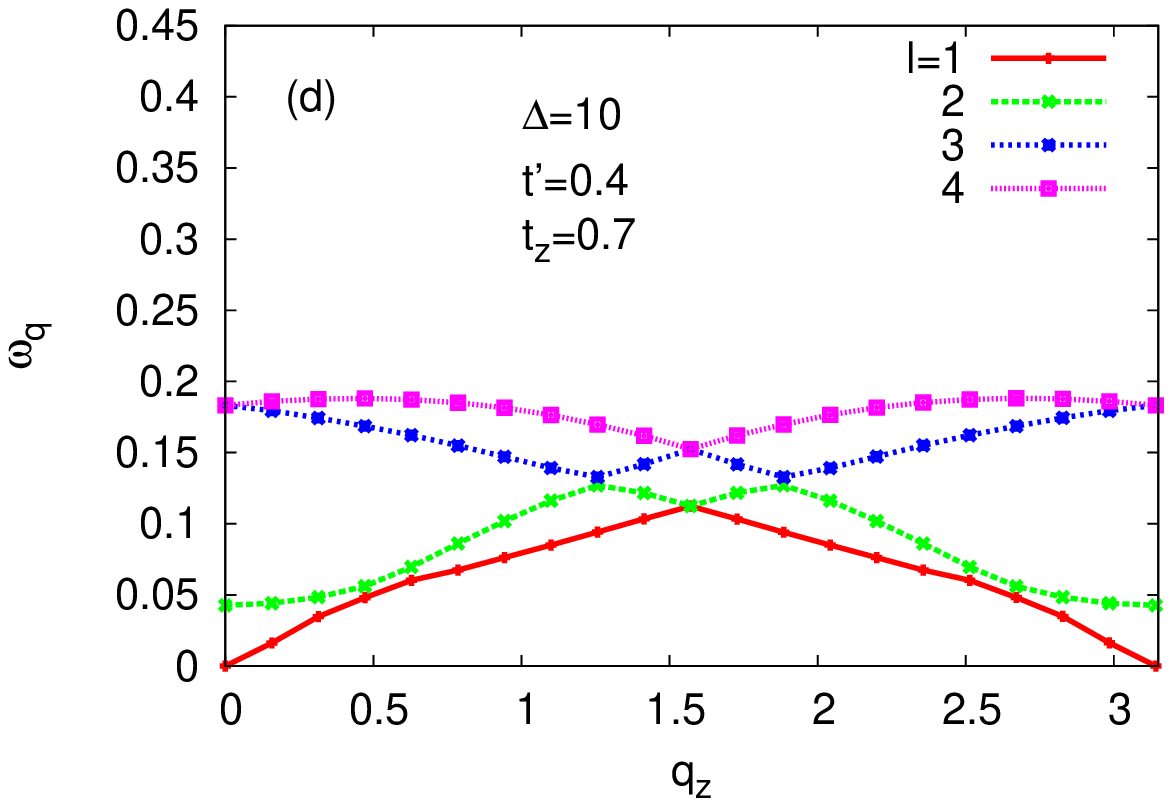,angle=0,width=70mm}
\ \\
\psfig{figure=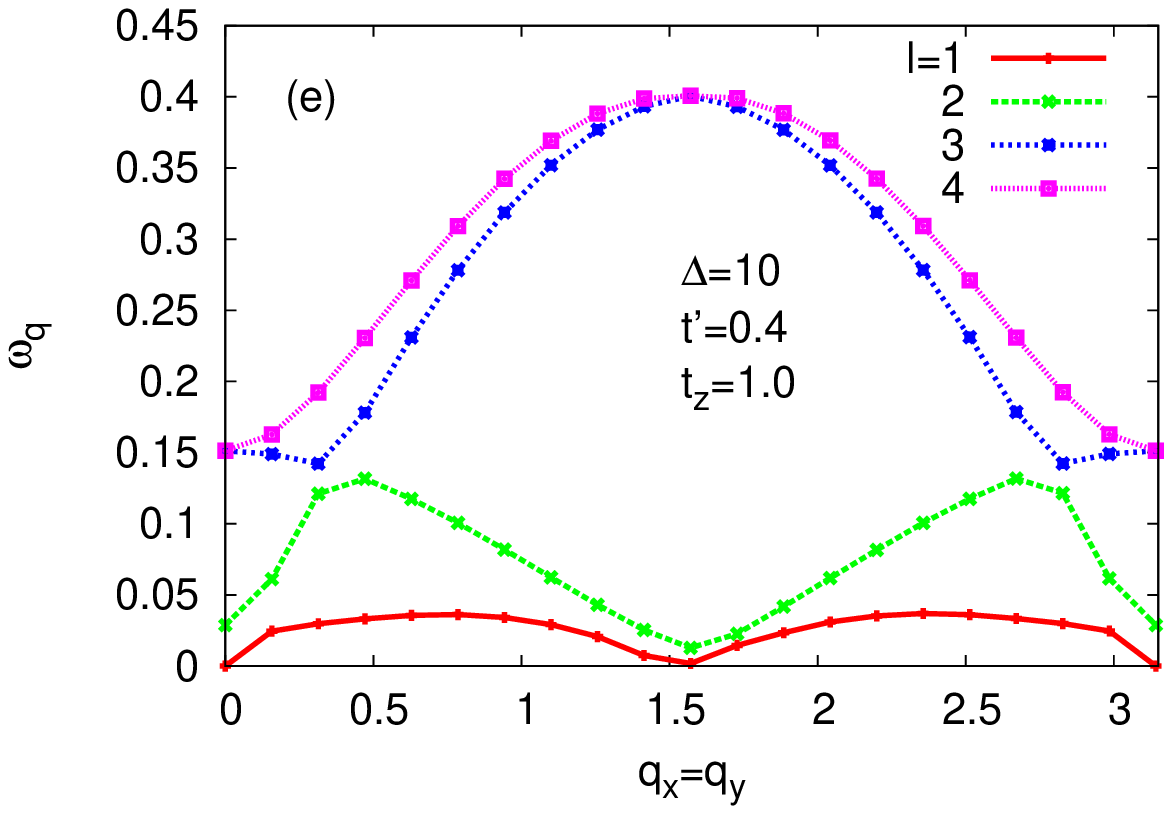,angle=0,width=70mm}
\psfig{figure=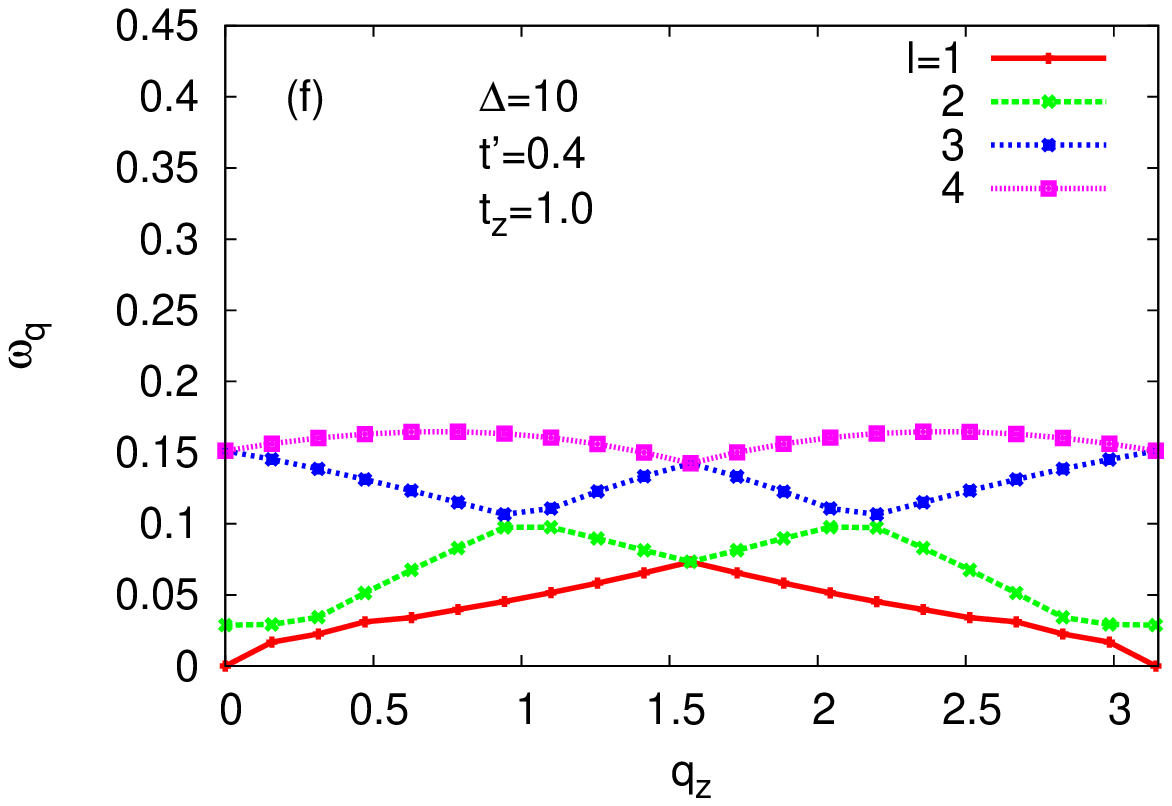,angle=0,width=70mm}
\caption{Calculated spin wave dispersion along planar ($q_x=q_y$) and perpendicular ($q_z$) directions. The layered AF subsystems on the two fcc sublattices are independent for $t_z=0$: (a) and (b), strongly coupled in the cubic case $t_z=1$: (e) and (f), and moderately coupled in the intermediate case $t_z = 0.7$: (c) and (d).}
\end{figure}

We will consider the planar ($q_z=0$, $q_x$ and $q_y$ finite) and perpendicular ($q_x$=$q_y$=0, $q_z$ finite) spin wave modes in this investigation, which will provide excitation energies corresponding to spin twisting within layers as well as how neighboring layers are magnetically coupled. We have mainly considered the case of positive $t'$ as it appears relevant for $\rm MnS_2$.

It is instructive to start with the limiting case $t_z$=0 where the two fcc sublattices get decoupled into simple layered antiferromagnetic subsystems with AF order in both planar and perpendicular directions. In this limit, the four spin wave branches (indicated by $l$=1-4) collapse into two [Figs. 3(a) and (b)], and the two Goldstone modes correspond to independent spin rotations in the two fcc sublattices. Further setting $t'$=0, all four branches become degenerate and match with the dispersion for the planar antiferromagnet. The calculated dispersion is of the form:
\begin{equation}
\omega_{\bf q} = (2+r) J \sqrt{1-\gamma_{\bf q} ^2}
\end{equation}
for a layered three-dimensional AF in the large $U$ limit, where 
\begin{equation}
\gamma_{\bf q} = (\cos q_x + \cos q_y + r\cos q_z)/(2+r)
\end{equation}
and $r$=$J'/J$=$(t'/t)^2$ is the ratio of the interlayer to planar spin couplings, with $J$=$4t^2/U$ and $J'$=$4t'^2/U$. Here the minor frustration effect due to small planar NNN hopping $t'$ has been neglected. From the above expression, the maximum ($q_x$=$q_y$=$\pi/2$) and zone boundary ($q_x$=$q_y$=$\pi$) energies for the planar mode are approximately $2J$ and $2J\sqrt{2}(t'/t)$, while at $q_z$=$\pi/2$ the perpendicular mode energy is $2J(t'/t)$=$2\sqrt{J J'}$. For $\Delta$=10 ($U$$\approx$20) and $t'/t$=0.4, we estimate these three energies as 0.4, 0.2, and 0.16, respectively, which validate the calculated results shown in Figs. 3(a) and (b). 

When $t_z$ is turned on, the layered AF subsystems on the two fcc sublattices get coupled and a pair of low-energy weakly dispersive branches emerge. The dispersion of the high-energy branches in Fig. 3(e) is similar as for uncoupled fcc sublattices ($t_z$=0). This reflects the inherent fcc lattice frustration, as discussed earlier. The low-energy branches, on the other hand, correspond to opposite spin twistings on the two fcc sublattices, resulting in healing of frustrated NN AF bonds and consequent lowering of energy. Only one Goldstone mode survives, and the other mode acquires a small energy gap at $q$=0 as seen in Fig. 3(e). This small energy gap reflects the effective magnetic coupling between the two fcc sublattices. Strong softening of the low-energy branches as $t_z$ approaches 1 (cubic case) highlights the fcc lattice frustration. The marginal stability of type-III order, as seen from the nearly vanishing energies at $q_x$=$q_y$=$\pi/2$ in Fig. 3(e), even in the strong coupling limit, possibly accounts for the rarity of this magnetic order in nature, and highlights the importance of magnetoelastic effect and weak magnetic anisotropy in the stabilization of type-III order in $\rm MnS_2$. 


\begin{figure}
\vspace*{0mm}
\hspace*{0mm}
\psfig{figure=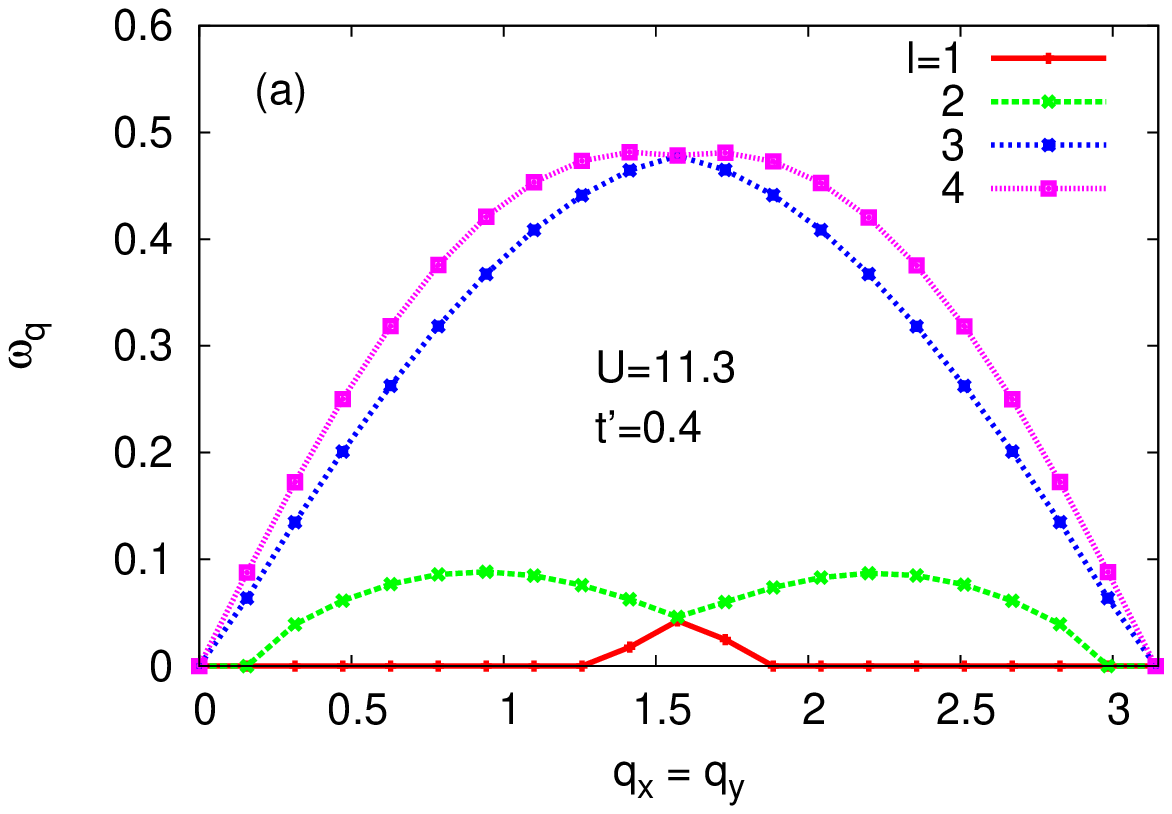,angle=0,width=80mm}
\psfig{figure=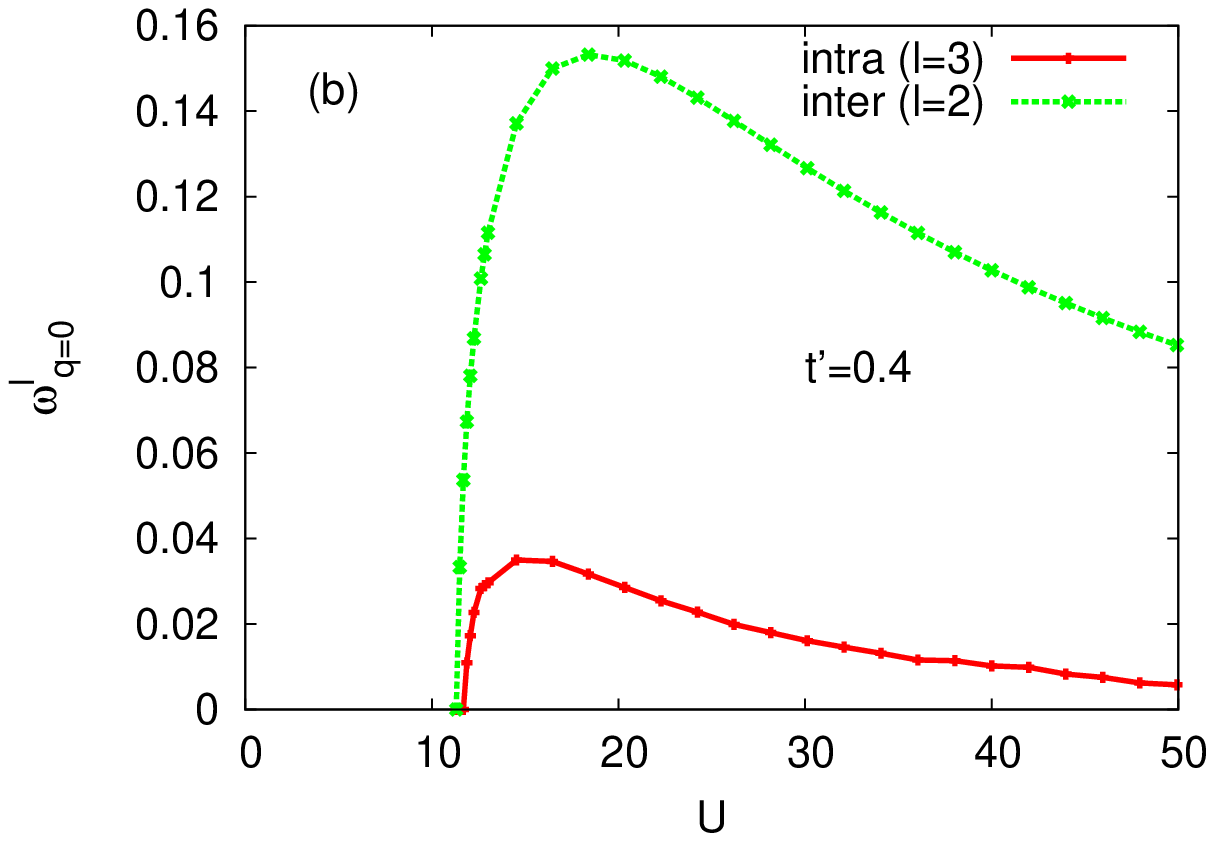,angle=0,width=80mm}
\ \\
\psfig{figure=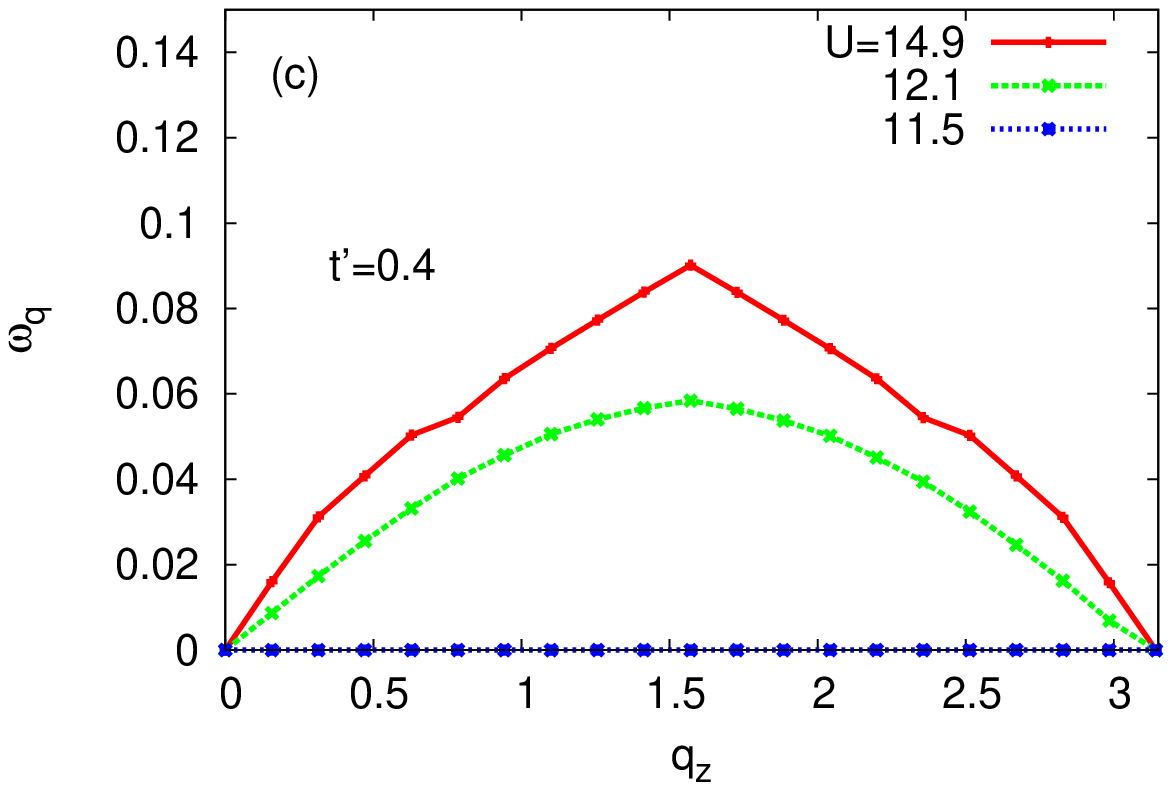,angle=0,width=80mm}
\psfig{figure=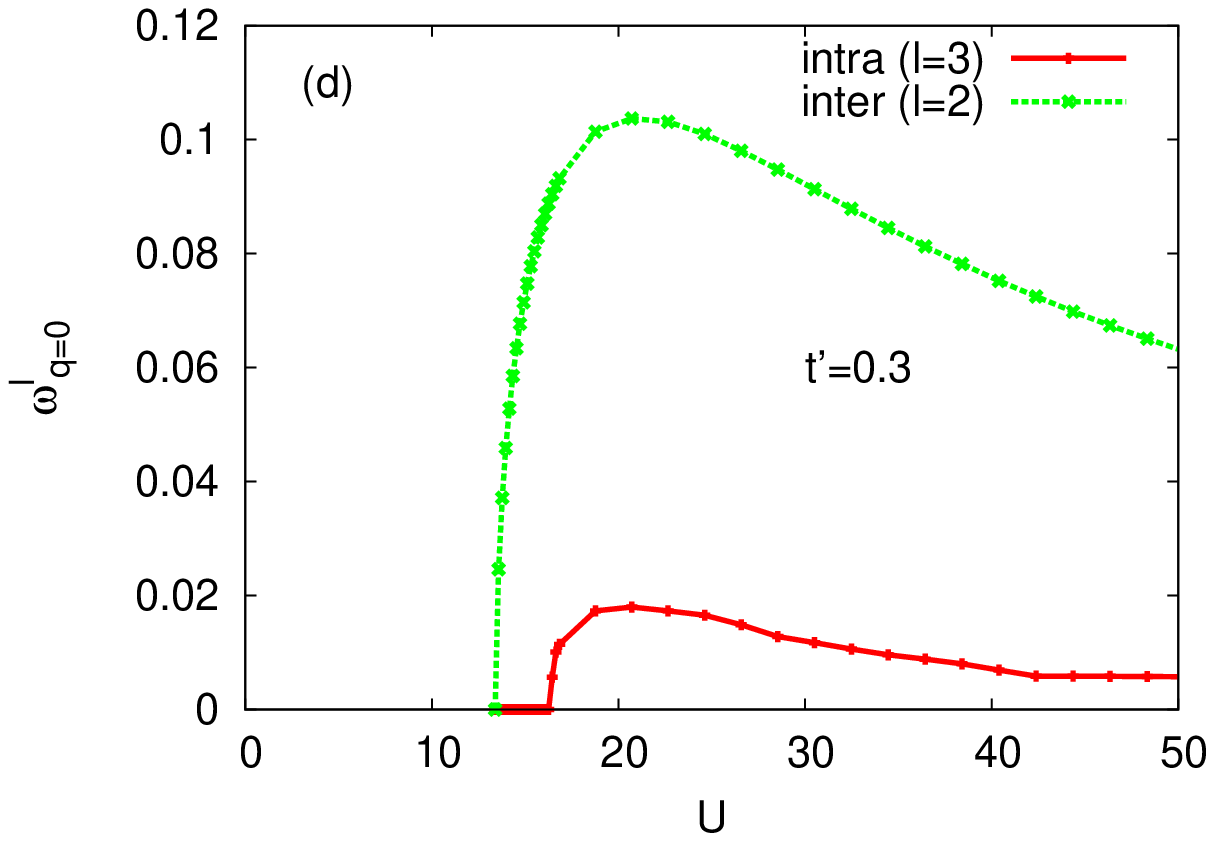,angle=0,width=80mm}
\caption{Finite-$U$-induced frustration effect on: (a) planar spin wave modes, (c) perpendicular mode (lowest-energy branch), and the characteristic energies $\omega_{q=0} ^l$ providing quantitative measure of the effective interlayer magnetic couplings (b) and (d).}
\end{figure}

With decreasing $U$, the low-energy branches undergo softening, eventually turning to negative-energy modes signalling instability of type-III order. Fig. 4(a) shows the planar dispersion at the $U$ value where the characteristic energy $\omega_{q=0} ^{l=3}$ just vanishes. Fig. 4(c) shows the softening of the lowest-energy perpendicular mode with decreasing $U$, the onset of negative-energy modes coinciding with the vanishing of the characteristic energy $\omega_{q=0} ^{l=2}$. These two characteristic energies provide quantitative measures of the effective interlayer magnetic couplings for same (intra) and different (inter) fcc sublattices, respectively. The reduction and eventual vanishing of these two energies with decreasing $U$ [Figs. 4(b) and (d)] highlights the finite-$U$-induced frustration effect in the fcc lattice, as explained below.

Effective hopping connections between parallel spins through pair of NN hoppings (indicated  in Fig. 1 as $t_z$) result in competing (i.e. antiferromagnetic) interactions at finite $U$ between 2nd neighbor (same layer) and 3rd neighbor (neighboring layers) spins. Unlike the weakly frustrated square-lattice AF involving weak NNN hopping terms $t'$, the finite-$U$-induced frustration effect in fcc lattice is quite significant as parallel spins are connected by NN hopping. 

The effective magnetic coupling between the two fcc sublattices is of particular interest. Within the localized spin model, NN interactions between spins on the two fcc sublattices are fully frustrated (Fig. 1), leading to degeneracy in the relative spin orientations. This degeneracy is, however, lifted at finite $U$ and the two fcc sublattices become effectively coupled. Of the two Goldstone modes ($q$=0) for $t_z$=0 corresponding to decoupled fcc sublattices, one mode acquires a small finite energy for finite $t_z$, and this energy $\omega_{q=0}^{l=2}$ provides a quantitative measure of the effective magnetic coupling between neighboring layers of different fcc sublattices. Energetically favorable magnetic coupling between the two fcc sublattices also confirms the stability of collinear type-III order. Figs. 4(b) and (d) show that this coupling vanishes at large $U$ corresponding to fully frustrated fcc sublattices.  

\begin{figure}
\psfig{figure=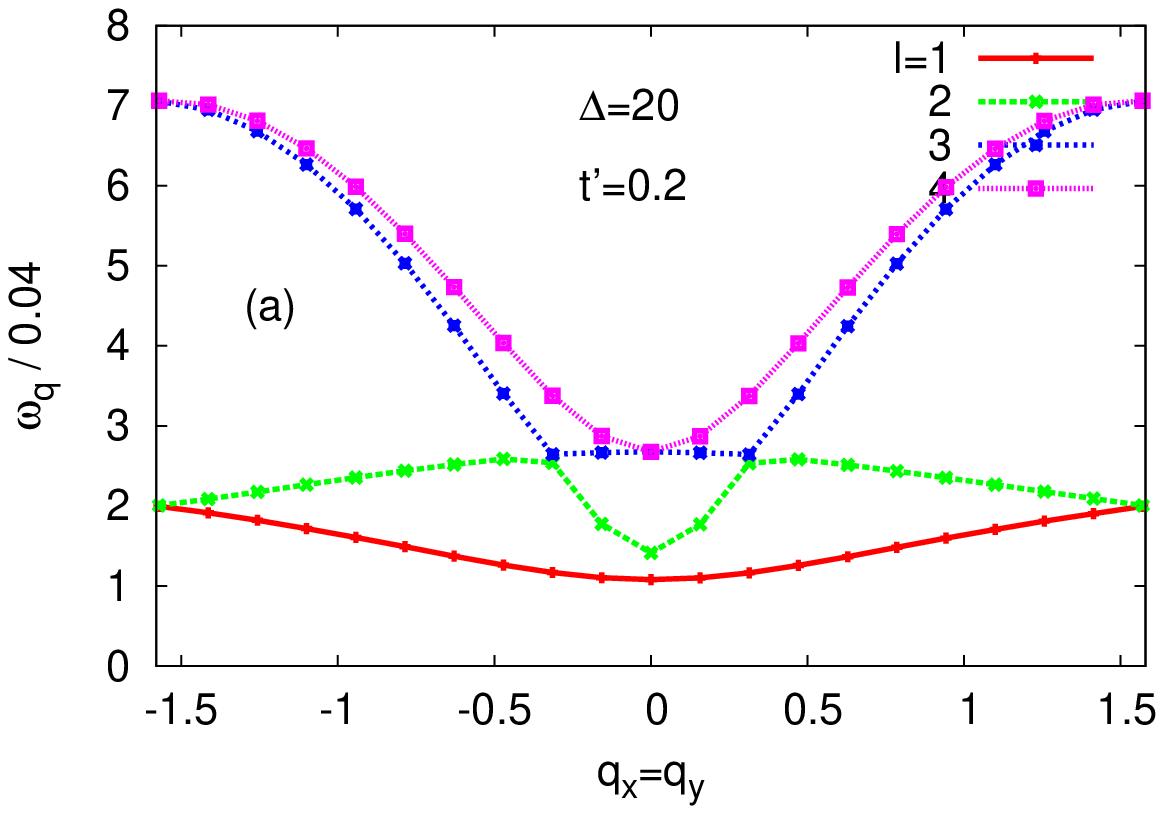,angle=0,width=80mm} \ \\
\psfig{figure=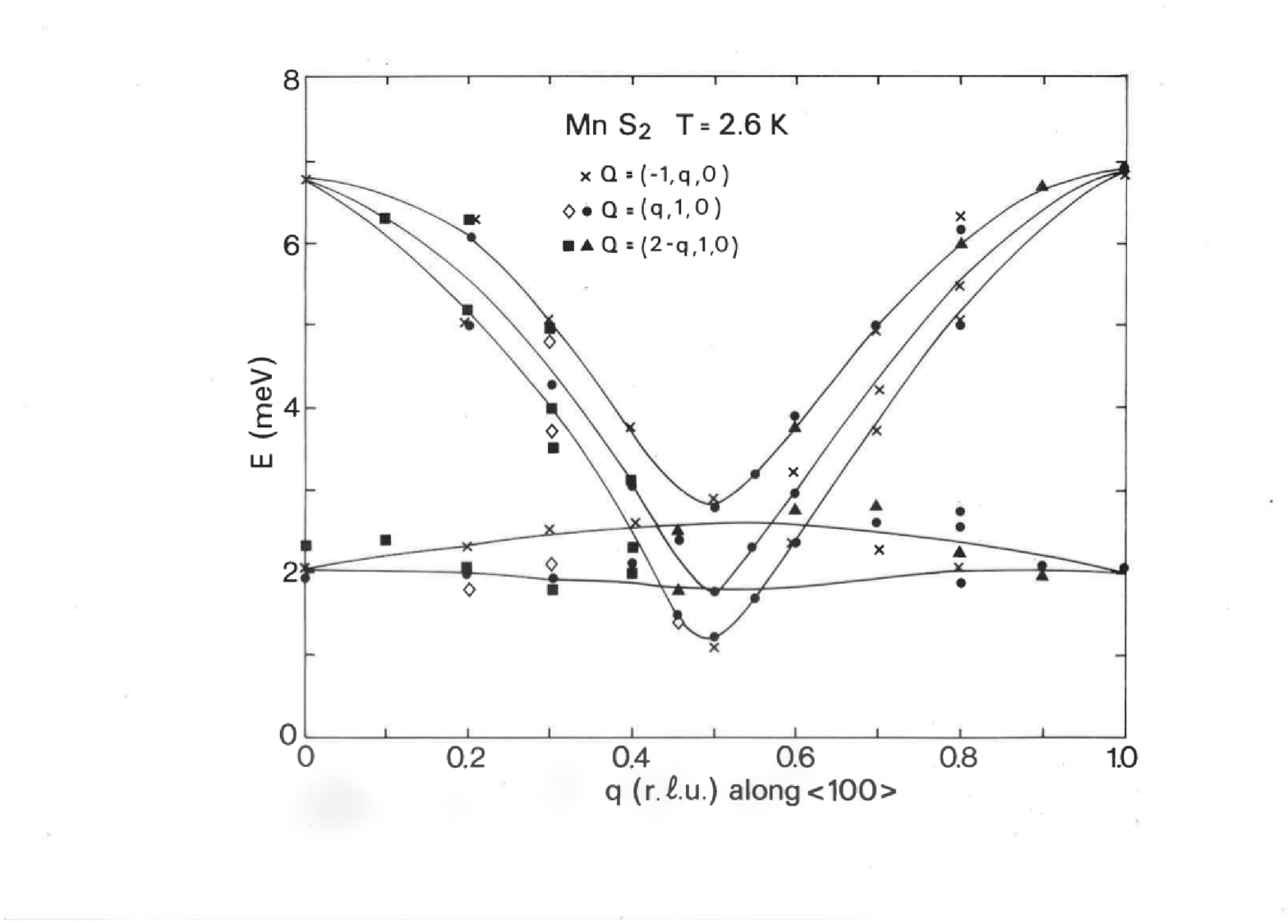,angle=-90,width=70mm}
\caption{(a) Calculated spin wave dispersion (planar mode) with small uniaxial anisotropy included. (b) Measured spin wave dispersion in $\rm MnS_2$ from inelastic neutron scattering experiments [12].}
\end{figure}

Figure 5(a) shows the calculated spin wave dispersion (all branches) for the planar mode in the momentum range 
$-\pi/2$$\le$$q$$\le$$\pi/2$. Here $U/t$$\approx$40, corresponding to the strong coupling limit, $t_z$=0.96, and a small uniaxial anisotropy term $ -\delta U \sum_i (S_i ^z)^2 $ was included $(\delta U/U$=$10^{-4})$ to stabilize type-III order and also phenomenologically account for the measured gap at the $\Gamma$ point. The calculated dispersion is in qualitative agreement with INS measurements\cite{tapan_mns2} of spin wave dispersion in $\rm MnS_2$ [Fig. 5(b)]. 


With decreasing $U$, enhancement of interlayer magnetic frustration results in magnetic instability when the spin wave energy turns negative. For fixed $t'$, type-III order is thus unstable below a critical interaction strength $U_c$. Similarly, for fixed $U$, decreasing $t'$ leads to the instability when the weak AF interlayer coupling due to $t'$ is unable to compete against the frustrating interlayer spin couplings generated by $t_z$. With increasing $t'$, the instability occurs at lower $U_c$ values where the finite-$U$-induced frustration is more effective, resulting in a characteristic negative slope of $U_c$ vs. $t'$. 

With increasing $t'$, a different kind of magnetic instability is obtained at $t'$$\approx$0.7 in the large $U$ limit involving competition between planar interactions. The instability expectedly shows up in the planar spin wave mode, as seen from emergence of negative energy modes at small $q$ [Fig. 6(a)]. This is the instability toward type-II order, and is related to the known instability in the planar antiferromagnet from $(\pi,\pi)$ to $(\pi,0)$ order as $t'/t \rightarrow 1/\sqrt{2}$. The perpendicular mode remains stable near this $t'$ value. For negative $t'$ also, the planar mode shows instability for $|t'|$ near $1/\sqrt{2}$ [Fig. 6(b)]. 

Some of our results are in agreement with the $n$=1 phase diagram obtained in Ref. [11] (where sign of $t'$ is reversed compared to our model). These include: i) negative slope of $U_c$ vs. $t'$, ii) the instability to type-II order 
at $|t'|$$\approx$0.7 in the strong coupling limit, and iii) the PM metal state for negative $t'$ (positive $t'$ in Ref. [11]) at lower $U$ values due to the strong frustration-induced band broadening and overlap of the two SDW bands (Fig. 2). However, our result is significantly different at lower $t'$ values (below 0.3), where we find sharp increase in $U_c$ below which finite-$U$-induced frustration destabilizes type-III order, as seen in Fig. 7. Also, we do not find the instability towards type-I order or the $(0,0,Q)$ spiral structure at lower $U$ values, as discussed below. 

\begin{figure}
\vspace*{0mm}
\hspace*{0mm}
\psfig{figure=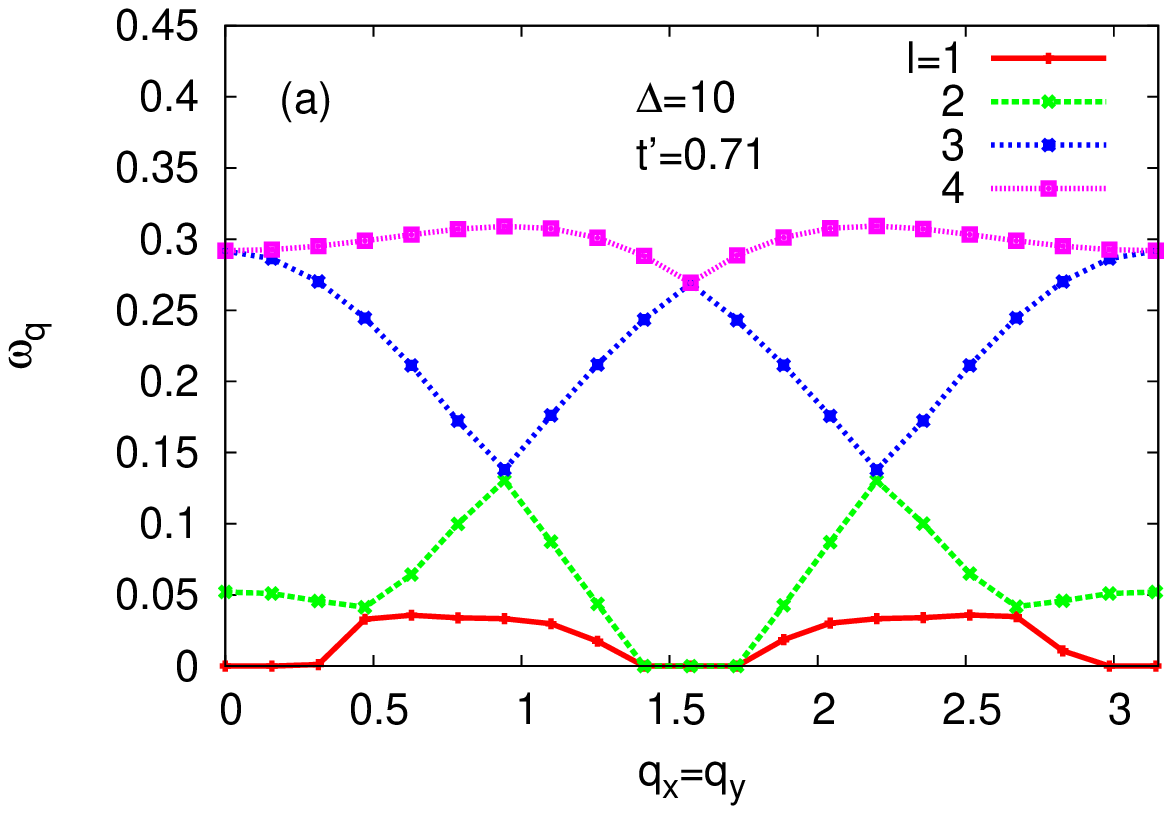,angle=0,width=80mm}
\psfig{figure=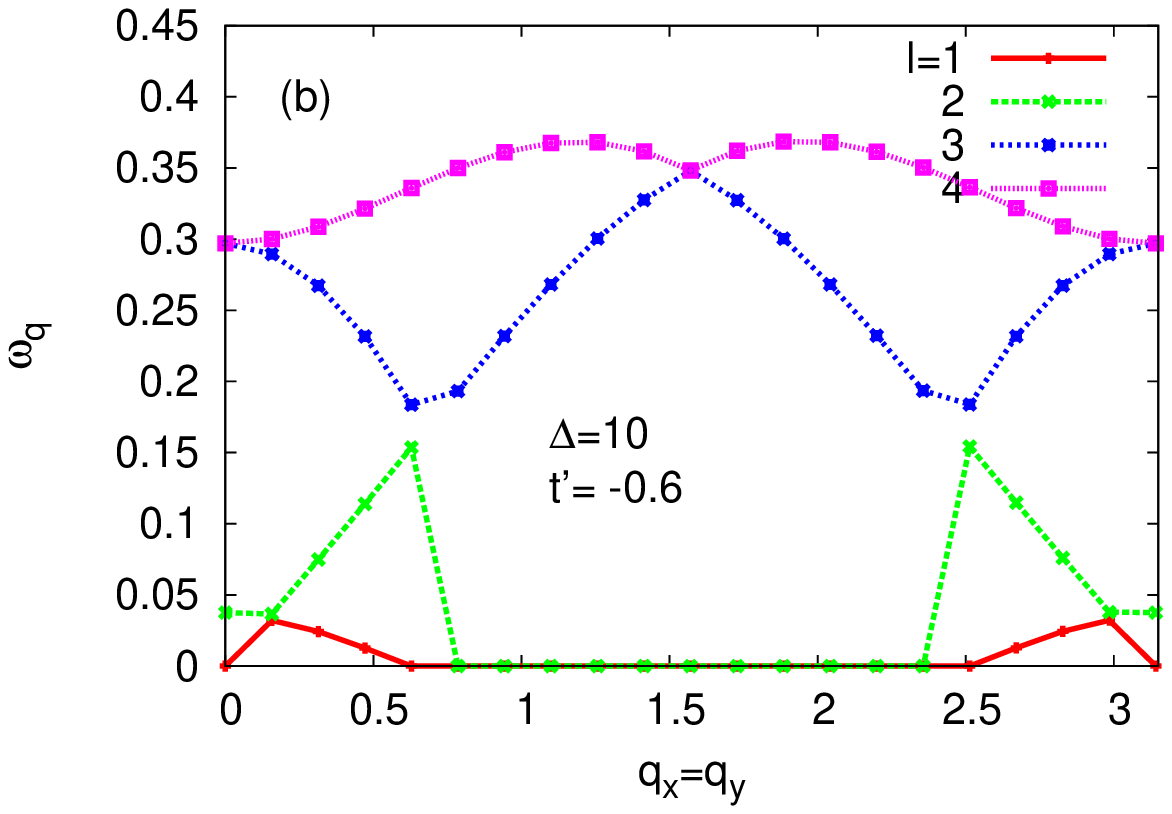,angle=0,width=80mm}
\caption{(a) Negative-energy modes at small $q$ for $t' \approx 1/\sqrt{2}$ signal long-wavelength instability towards type-II order which has $(\pi,0)$ instead of $(\pi,\pi)$ planar magnetic order. (b) For negative $t'$, the instability towards type-II order at $|t'| \approx 1/\sqrt{2}$ extends over a broad momentum range.}
\end{figure}

\begin{figure}
\vspace*{0mm}
\hspace*{0mm}
\psfig{figure=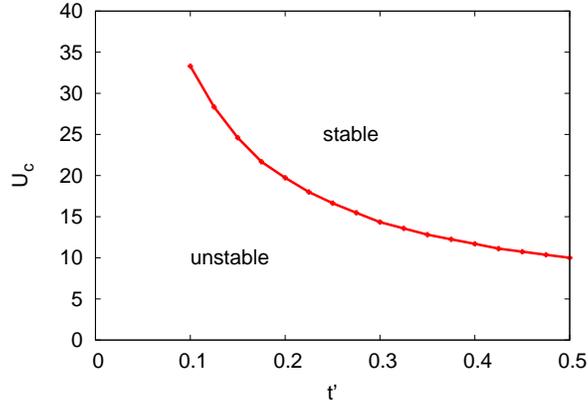,angle=0,width=80mm}
\caption{Critical interaction $U_c$ vs. $t'$ for stability of type-III order, based on spin wave stability analysis using the lowest-energy branch of the perpendicular mode [Fig. 4(c)]. Due to weaker AF interlayer coupling at lower $t'$ values, smaller magnitude of finite-$U$-induced frustration is sufficient for destabilization, which accounts for the sharp increase in $U_c$.}
\end{figure}

Competition between effective interlayer couplings for different and same fcc sublattices would result in the spiral structure along $z$ direction. However, we do not find this $(0,0,Q)$ spiral structure instability which would be signalled by negative energy modes at small but finite $q_z$. Instead, with decreasing $U$, we find that $\omega_{q=0}^{l=2}$ decreases to zero and turns negative, resulting in negative energy modes for all $q_z$. This implies that the coupling between different fcc sublattices is turning negative, signalling instability towards non-collinear order involving relative spin twisting between the two fcc sublattices. We have also examined spin waves in the type-I magnetic structure in the region marked ``unstable" for type-III order (Fig. 7). We find type-I order to be also unstable (Fig. 8), indicating that instability of type-III order as inferred from the perpendicular spin wave mode is not towards type-I order or the $(0,0,Q)$ spiral structure but rather towards non-collinear order.  

\begin{figure}
\vspace*{0mm}
\hspace*{0mm}
\psfig{figure=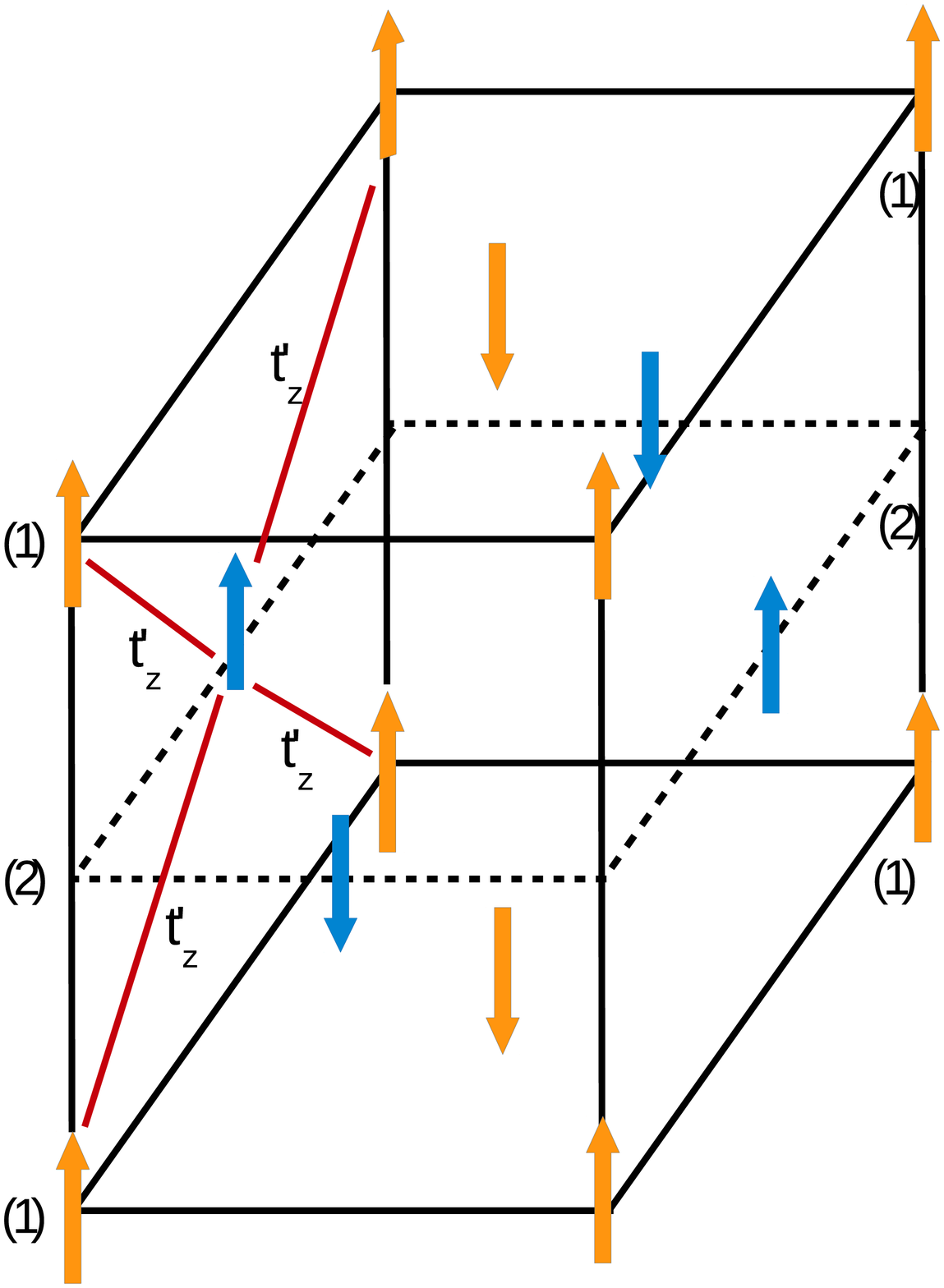,angle=0,width=50mm}
\psfig{figure=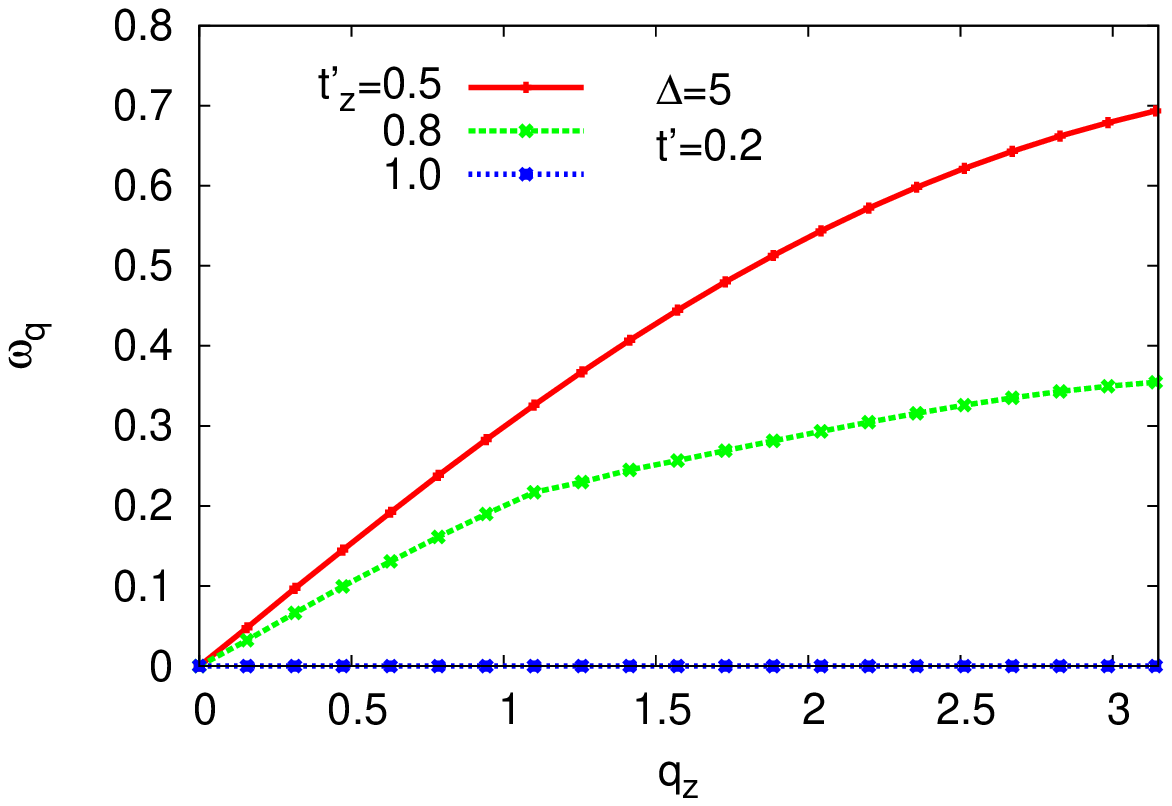,angle=0,width=80mm}
\caption{(a) Type-I order on the fcc lattice. (b) Perpendicular mode spin wave energy (lowest-energy branch) in the type-I ordered AF state, showing that type-I order is stable when the frustrating hopping term $t_z '$ is reduced but becomes unstable as $t_z ' \rightarrow 1$ in the cubic limit.} 
\end{figure}

Finally, we consider the factors qualitatively affecting the particle-hole gap for a multi-band system with crystal-field (CF) splitting and Hund's-rule coupling term included. Fig. 9 schematically shows CF split lower and upper SDW sub-bands, with the corresponding majority spins ($\uparrow$ and $\downarrow$) on the A sublattice indicated. The effective particle-hole gap is between the upper and lower SDW sub-bands corresponding to the lower and upper CF levels, respectively. With increasing $\Delta E_{\rm CF}$ due to pressure, overlap of these two SDW sub-bands and filling of the upper sub-band (spin $\downarrow$) at the expense of lower sub-band (spin $\uparrow$) results in a pressure-induced metal-insulator transition accompanied with high-spin to low-spin transition. Strongly reduced $\Delta E_{\rm SDW} - W$ due to increased bandwidth $W$ in the frustrated fcc lattice, together with exceptionally large increase in $\Delta E_{\rm CF}$ with pressure (inferred from the observed lattice-parameter reduction), are the likely favourable factors for pressure-induced metal-insulator transition in compounds such as $\rm Mn Te_2$. 

\begin{figure}
\vspace*{0mm}
\hspace*{0mm}
\psfig{figure=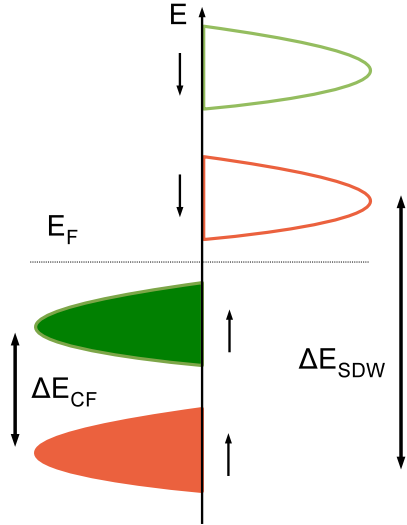,angle=0,width=60mm}
\caption{Schematic diagram showing crystal-field split lower and upper sub-bands in the SDW state of a half-filled multi-orbital model. The nominal particle-hole gap $\Delta E_{\rm ph} = \Delta E_{\rm SDW} - \Delta E_{CF} - W$ decreases with increasing crystal field splitting, illustrating the mechanism of pressure-induced metal-insulator transition with the onset of band overlap.} 
\end{figure}

\section{Summary and Discussion}

Spin waves in the type-III ordered AF state of the $t$-$t'$ Hubbard model on the fcc lattice were investigated. A composite four-layer, two-sublattice basis was employed corresponding to the $\alpha \alpha' \beta \beta' \alpha ...$ sequence of layers, and NN hopping terms between the two fcc sublattices were used as control to highlight magnetic frustration in the fcc lattice. Clearly illustrated by the reduction and eventual vanishing of the effective interlayer magnetic couplings with decreasing $U$, strong finite-$U$-induced competing interactions result in significant spin wave softening, besides the usual geometric frustration effect. Calculated spin wave dispersion with a weak magnetic anisotropy term included for stabilization was found to be in qualitative agreement with the measured dispersion in $\rm Mn S_2$ obtained from inelastic neutron scattering experiments.

The delicate energy balance between competing magnetic interactions results in extreme sensitivity to Hamiltonian parameters, leading to sharp instabilities as inferred from spin wave energies turning negative. While instabilities towards type-I order and $(0,0,Q)$ spiral structure were not observed, instability towards non-collinear order was inferred from the perpendicular mode, indicating relative spin twisting between different fcc sublattices due to vanishing of corresponding interlayer magnetic coupling. The planar mode also showed instability near $q_x$=$q_y$=$\pi/2$ which sets in at slightly higher $U$ values. The instability to type-II order near $t'$$\approx$1/$\sqrt{2}$ corresponds to the known instability in the frustrated planar AF from $(\pi,\pi)$ to $(\pi,0)$ magnetic order. 


The strong frustration effects manifested in the fcc lattice AF provide understanding of the unusual magnetic properties of the fcc-structure compounds ($\rm Mn S_2$, $\rm Mn Se_2$, and $\rm Mn Te_2$), such as the critical role of magnetoelastic effect and weak magnetic anisotropy in stabilizing type-III order and the weakly dispersive spin wave branch observed in $\rm Mn S_2$. A likely scenario for the first order magnetic transition observed near $T_N$ is that loss of inter-layer spin correlations near $T_{\rm N}$ due to thermal spin disordering suppresses the magneto-elastic effect and dipolar energy gain, which further enhances thermal spin fluctuations, resulting in a runaway effect which causes the first order magnetic transition.

Furthermore, the reduced SDW band gap due to strong frustration-induced band broadening and self-energy corrections renders the frustrated fcc lattice AF particularly susceptible to vanishing band gap with decreasing $U/t$. The above band picture of metal-insulator transition due to band overlap captures the essential feature in the realistic multi-band scenario involving interplay between Hund's-rule coupling and crystal-field splitting. Increasing crystal-field splitting with applied pressure reduces the energy gap between the highest occupied and the lowest unoccupied crystal-field sub-bands, with the pressure-induced metal-insulator transition corresponding to the onset of band overlap. This may be relevant to the pressure-induced high-spin to low-spin magnetic transition observed in $\rm Mn Te_2$ accompanied with changes in transport behavior suggestive of metal-insulator transition. 

\section*{Acknowledgement}
Helpful discussions with Mike Zhitomirsky are gratefully acknowledged.


\section*{References}

\begin{thebibliography}{99}

\bibitem{lacroix_2011} C. Lacroix, P. Mendels, and F. Mila, 
{\em Introduction to Frustrated Magnetism}, Springer, Berlin (2011).

\bibitem{hastings_1959} J. M. Hastings, N. Elliott, and L. M. Corliss,
Phys. Rev. {\bf 115}, 13 (1959).

\bibitem{bloch_1974} D. Bloch, R. Maury, C. Vetter, and W. B. Yelon, 
Phys. Lett. {\bf 49A}, 354 (1974).

\bibitem{capone_2009} M. Capone, M. Fabrizio, C. Castellani, and E. Tosatti, 
Rev. Mod. Phys. {\bf 81}, 943 (2009).

\bibitem{ganin_2010} A. Y. Ganin {\em et al.}, Nature {\bf 466}, 221 (2010).

\bibitem{guiot_2013} V. Guiot, L. Cario, E. Janod, B. Corraze, V. Ta Phuoc, M. Rozenberg, P. Stoliar, T. Cren, and D. Roditchev, Nature Communications {\bf 4}, 1722 (2013). 

\bibitem{aczel_2013} A. A. Aczel, D. E. Bugaris, L. Li, J.-Q. Yan, C. de la Cruz, H.-C. zur Loye, 
and S. E. Nagler, Phys. Rev. B {\bf 87}, 014435 (2013).

\bibitem{as_2005} A. Singh, Phys. Rev. {\bf 71}, 214406 (2005).

\bibitem{coldea_2001} R. Coldea, S.M. Hayden, G. Aeppli, T.G. Perring, C.D. Frost, T.E.
Mason, S.-W. Cheong, and Z. Fisk, Phys. Rev. Lett. {\bf 86}, 5377 (2001), and references therein. 

\bibitem{asingh_2004+5} P. Srivastava and A. Singh, 
Phys. Rev. B {\bf 70}, 115103 (2004); Phys. Rev. B {\bf 72}, 224409 (2005).

\bibitem{timirgazin_2016} M. A. Timirgazin, P. A. Igoshev, A. K. Arzhnikov, and V. Yu. Irkhin, 
arxiv: 1603.03242 (2016). 

\bibitem{tapan_mns2} T. Chattopadhyay, P. Burlet, L. P. Regnault and J. Rossat-Mignod, 
Physica B {\bf 156 \& 157}, 241 (1989).

\bibitem{tapan_1991} T. Chattopadhyay, P. Burlet and P. J. Brown,
J. Phys.: Condens. Matter {\bf 3}, 5555 (1991).

\bibitem{hastings_1976} J. M. Hastings and L. M. Corliss, 
Phys. Rev. B {\bf 14}, 1995 (1976).

\bibitem{tapan_1984} T. Chattopadhyay, H. G. von Schnering and H. A. Graf, 
Solid State Commun. {\bf 50}, 865 (1984).

\bibitem{westrum_1970} E, F. Westrum Jr. and F. Grenvold, 
J. Chem. Phys. {\bf 52}, 3870 (1970).

\bibitem{kimber_2015} S. A. J. Kimber and T. Chatterji,
J. Phys.: Condens. Matter {\bf 27}, 226003 (2015).

\bibitem{kimber_2014} S. A. J. Kimber, A. Salamat, S. R. Evans, H. O. Jeschke, K. Muthukumar,
M. Tomi\'{c}, F. Salvat-Pujol, R. Valentí, M. V. Kaisheva, I. Zizak, and T. Chatterji, 
PNAS {\bf 111}, 5106 (2014). 

\bibitem{tapan_2015} T. Chatterji, A. M. dos Santos, J. J. Molaison, T. C. Hansen, S. Klotz, M. Tucker, 
K. Samanta, and T. Saha-Dasgupta, Phys. Rev. B {\bf 91}, 104412 (2015).

\bibitem{mita_2008} Y. Mita, Y. Ishida, M. Kobayashi and S. Endo
Acta Phys. Pol. A, {\bf 113}, 617 (2008).

\bibitem{gvozdikova_2005} M. V. Gvozdikova and M. E. Zhitomirsky,
arXiv: 0502255 (2005). 

\bibitem{qmc1} S. A. Brazovskii, I. E. Dzyaloshinskii, and B. G. Kukharenko, 
Zh. Eksp. Teor. Fiz. {\bf 70}, 2257 (1976) [Sov. Phys. JETP {\bf 43}, 1178 (1976)].

\bibitem{qmc2} D. Mukamel and S. Krinsky, Phys. Rev. B {\bf 13}, 5078 (1976).

\bibitem{henley_1987} C. L. Henley, J. Appl. Phys. {\bf 61}, 3962 (1987).

\bibitem{canals_2004} B. Canals and M. E. Zhitomirsky, 
J. Phys.: Condens. Matter {\bf 16}, S759 (2004).  

\end{thebibliography}
\end{document}